\documentclass[a4paper,11pt]{article}
\pdfoutput=1
\usepackage{jheppub1}
\usepackage{amsmath,amssymb,color,graphics,epsfig}
\usepackage{hyperref}
\usepackage{tikz}
\usetikzlibrary{shapes.geometric, arrows, decorations.pathreplacing}

\usepackage{amsfonts}
\usepackage[all]{xy}

\newcommand{\be}{\begin{equation}}
	\newcommand{\ee}{\end{equation}}
\newcommand{\bea}{\setlength\arraycolsep{2pt} \begin{eqnarray}}
	\newcommand{\eea}{\end{eqnarray}}
\newcommand{\nn}{\nonumber}

\def\fft#1#2{{\frac{#1}{#2}}}

\def\0{{\sst{(0)}}}
\def\1{{\sst{(1)}}}
\def\2{{\sst{(2)}}}
\def\3{{\sst{(3)}}}
\def\4{{\sst{(4)}}}
\def\5{{\sst{(5)}}}
\def\6{{\sst{(6)}}}
\def\7{{\sst{(7)}}}
\def\8{{\sst{(8)}}}
\def\sst#1{{\scriptscriptstyle #1}}

\def\pd{\partial}
\def\tg{\tilde{g}}
\def\d{\mathrm{d}}
\def\f{\mathfrak{f}}
\def\tc{\tilde{c}}

\def\e {\mathrm{e}}

\begin{document}
\title{From Specific Heat Bounds to Penrose-like Inequalities: A Unified Thermodynamic Perspective}

\author{Kai-Peng Lu, H. L\"{u} and Run-Qiu Yang$^{1}$}
\affiliation{Center for Joint Quantum Studies, Department of Physics, School of Science, Tianjin University, Yaguan Road 135, Jinnan District, 300350 Tianjin, P.~R.~China}
\emailAdd{kaipenglu@tju.edu.cn}
\emailAdd{mrhonglu@gmail.com}
\emailAdd{aqiu@tju.edu.cn}
\footnotetext[1]{The corresponding author.}
\abstract{Under explicit and physically motivated assumptions, we identify a local-to-global thermodynamic mechanism that organizes several previously disconnected inequalities on black holes. We first focus on asymptotically flat, static, and spherically-symmetric black holes, and provide a unified thermodynamic perspective on various Penrose-like inequalities. We also discuss the generalizations in asymptotically Schwarzschild  anti-de Sitter black holes/branes. In holographic settings, these bounds imply monotonic behavior of normalized thermodynamic quantities in the dual field theory. Notably, one central integrated inequality can be established directly under the NEC in general static planar symmetric asymptotically Schwarzschild anti-de Sitter black branes. Our purpose is not to claim a fully general proof of the proposed local bounds, but to expose a  unified structural perspective connecting local thermodynamic constraints to global geometric inequalities. Establishing the local bounds beyond the classes considered here remains an open problem. 
}

\maketitle

\section{Introduction} \label{sec:introduction}

In general relativity, black holes are geometric solutions characterized by a small set of conserved quantities \cite{Israel:1967wq,Carter:1971zc,Hawking:1971vc}, yet semiclassical analyses reveal that they operate as thermodynamic systems \cite{Hawking:1971tu,Hawking:1975vcx,Bekenstein:1973ur}, whose physical evolution must comply with the laws of thermodynamics \cite{Hawking:1976de,Bardeen:1973gs}. This implies that the macroscopic geometric parameters of black holes are constrained by thermodynamic laws. Consequently, establishing algebraic relations among these physical variables provides a concrete criterion to delineate the permissible configurations of black hole spacetimes.

For asymptotically flat black holes, the most prominent example of such a relation is the Penrose inequality \cite{Penrose:1973um,Bray:2003ns,Mars:2009cj,Ben-Dov:2004lmn}. Expressed in terms of the total Arnowitt-Deser-Misner (ADM) mass $M$ and the entropy $S$, the classical inequality takes the form\footnote{In this paper we take the conventions: $G=c=\hbar=k_B=1$. }
\be
2M \ge \sqrt{S/\pi} \,. \label{Penrose0}
\ee
This is the form in 4-dimensional asymptotically flat spacetimes. We can rewrite it as
\be
S(M)\leq S_{\text{Sch}}(M) \,. \label{Penrose}
\ee
Here $S(M)$ is the entropy of a black hole with mass $M$ and $S_{\text{Sch}}(M)$ is the entropy of a Schwarzschild black hole with the same mass $M$. This form can be generalized to arbitrary dimensions of asymptotically flat or anti-de Sitter (AdS) spacetimes. This inequality could be interpreted physically: the Schwarzschild black hole will have maximal entropy among all black holes of same energy. Rigorous mathematical proofs of \eqref{Penrose} in some special cases via various energy conditions can be found in \cite{Bray:1999eow,Huisken:2001,Hayward:1994bu,Xiao:2022obq}.  A general proof for inequality~\eqref{Penrose} is still open, see Ref.~\cite{Mars:2009cj} for a review on this topic. Note that both mass $M$ and entropy $S$ are geometrical quantities in that they can be derived directly from the metric, without having to know the details of the matter fields. Recent investigations have expanded upon the Penrose inequality by explicitly incorporating another geometric quantity, namely the black hole temperature $T$. This leads to two notable variants. The first is a ``relaxed'' version \cite{Yang:2022yye} established for spherically-symmetric and static black holes relying on the null energy condition (NEC),
\be
3M \ge \sqrt{S/\pi} + 2TS \,. \label{Penrose-relax}
\ee
It is easy to see that \eqref{Penrose-relax} becomes less restrictive for the extremal and near-extremal black holes. To provide a stricter constraint, a ``tightened'' Penrose-like inequality was subsequently proposed \cite{Khodabakhshi:2022jot}, reading
\be
M \ge \sqrt{S/\pi} - 2TS \,. \label{Penrose-tighten}
\ee
The proof of this tightened bound has been given in static spherically symmetric case and necessitates a combination of the NEC and the trace energy condition (TEC). Physically, \eqref{Penrose-tighten} provides a more accurate lower bound for the mass of extremal configurations and is saturated by both the Schwarzschild and Reissner-Nordstr\"om (RN) black holes. In addition, the direct addition of these two variants recovers the classical bound \eqref{Penrose}.

While the aforementioned Penrose-like inequalities each possess distinct physical motivations and require different energy conditions for their proofs, they share a structural consistency that hints at a potential unified thermodynamic origin. To explore this possibility, it is instructive to consider the black hole state in a fixed-charge ensemble, where the conserved charges $Q_i$ are held constant, and thermodynamic variables such as the mass $M$ and temperature $T$ are treated as functions of the entropy $S$ alone. From this perspective, the validity of a global algebraic bound of the form $g(S) \ge 0$ can be investigated through the lens of differential analysis. Specifically, if the derivative $g'(S)$ exhibits a definite sign, the proof of $g(S) \ge 0$ is then reduced to evaluating the function at a physical boundary, such as the Schwarzschild limit. Since $g'(S)$ involves the specific heat $C_{Q_i} = T(\pd S/\pd T)_{Q_i}$, this differential perspective suggests that the various Penrose-like inequalities may originate from an underlying constraint on black hole specific heat.

The physical relevance of constraining the specific heat has recently been highlighted in the context of the weak cosmic censorship conjecture (WCCC) \cite{Lu:2025iky}. Building upon the perturbative approach in gedanken experiments proposed by Sorce and Wald for the Kerr-Newman black hole \cite{Sorce:2017dst,Wald:2018xxi}, recent studies \cite{Wu:2024ucf,Lu:2025ntu} demonstrated that the validity of the WCCC for extremal and near-extremal black holes hinges entirely on the sign of a single quantity at the zero-temperature limit $W = (\pd S/\pd T)_{Q_i; T=0}$. Generalizing this criterion to finite temperatures naturally invokes the specific heat $C_{Q_i}$. However, because $C_{Q_i}$ diverges at points of thermodynamic phase transitions \cite{Davies:1977bgr,Hawking:1982dh}, it is mathematically more robust to investigate its inverse,
\be
U \equiv \bigg(\fft{\pd T}{\pd S} \bigg)_{Q_i} = T/C_{Q_i} \,. \label{U}
\ee
Consequently, an inequality bounding $U$ was established for spherically-symmetric and static black holes in asymptotically flat spacetime \cite{Lu:2025iky},
\be
-\fft{T}{2S} \le U \le \fft1{ 4\sqrt{\pi S^3}}\,,   \label{asy-heatbound-4D}
\ee
where the lower and upper bounds can be ensured by SEC and DEC respectively.
In this work, we demonstrate that this specific heat bound provides the missing structural ingredient: it enforces the monotonicity of a class of auxiliary functions, from which the Penrose-like inequalities follow in a unified manner. Specifically, the lower bound in \eqref{asy-heatbound-4D} ensures the monotonicity required to directly derive the relaxed Penrose inequality \eqref{Penrose-relax}.

Noting that the tightened Penrose inequality \eqref{Penrose-tighten} is saturated by both Schwarzschild and RN black holes, we expect that a corresponding local thermodynamic bound should be compatible with these saturations. However, the upper bound in \eqref{asy-heatbound-4D} is saturated only by the extremal RN black holes \cite{Lu:2025iky}, indicating the bound is not restrictive enough. We therefore propose a refinement that matches the saturation for all RN black holes, which leads to a more restrictive upper bound:
\be
-\fft{T}{2S} \le U \le \fft1{ 4\sqrt{\pi S^3}} -\fft{3T}{2S} \,. \label{asy-golden-4D}
\ee
This refined upper bound is simultaneously saturated by both the Schwarzschild and generic RN black holes. Indeed, we shall show that it leads to \eqref{Penrose-tighten} as a direct consequence. This suggests that the refined bound \eqref{asy-golden-4D} provides a unified thermodynamic framework underlying these geometric inequalities.

Beyond asymptotically flat spacetimes, the generalization to anti-de Sitter (AdS) backgrounds introduces a nontrivial vacuum energy due to the negative cosmological constant, rendering the naive extension of \eqref{asy-golden-4D} ambiguous. We demonstrate that this issue can be resolved by subtracting a suitable background contribution following the method of ``holographic renormalization'' scheme. While conceptually sharing the same physical motivation as the sophisticated procedure of holographic renormalization \cite{Balasubramanian:1999re, deHaro:2000vlm,Bianchi:2001kw,Skenderis:2002wp}, this direct reference background subtraction \cite{Witten:1998qj} isolates the finite thermodynamic contributions of the physical matter fields. The resulting renormalized inequality remains valid under appropriate energy conditions. Moreover, for planar AdS black holes, these bounds organize into an interesting hierarchy controlled by bulk energy conditions, revealing a structured interplay between the null, weak, strong, and dominant energy conditions.

Furthermore, this local thermodynamic constraint on $U$ can be generalized to arbitrary dimensional AdS spacetimes with planar symmetry. A central outcome of our analysis is that the renormalized specific heat bound admits a natural integrated form
\be
d(E - E_0) \ge (d-1) TS \,, \label{energy bound}
\ee
where $E_0$ denotes the reference energy in the vacuum case and $d$ is the spacetime dimension of the dual conformal field theory (CFT). This inequality is not an independent {\it ad hoc} assumption, but rather a direct consequence of the bounds on $U$. Although the local specific-heat bounds are established only under simplifying assumptions, one of their central integrated consequences, Eq.~\eqref{energy bound}, can be proven directly under substantially weaker assumptions (see Appendix~\ref{adsets1}).

Through the anti-de Sitter/conformal field theory (AdS/CFT) correspondence \cite{Witten:1998qj,Maldacena:1997re,Gubser:1998bc,Witten:1998zw}, this bound has direct implications for the dual CFT. Specifically, we will show later that it enforces the monotonic growth of a suitably normalized free energy density $\f$ with respect to the temperature. In quantum field theory, monotonic quantities often play a central role in characterizing the irreversible loss of effective degrees of freedom along renormalization group (RG) flows. Representative examples include Zamolodchikov's $c$-theorem in two dimensions \cite{Zamolodchikov:1986gt}, the $a$-theorem in four dimensions \cite{Cardy:1988cwa, Komargodski:2011vj}, and the $F$-theorem in three dimensions \cite{Jafferis:2011zi, Casini:2012ei}. In holographic systems at finite temperature, the temperature itself naturally introduces an energy scale, suggesting that thermodynamic quantities may also encode aspects of RG monotonicity. Therefore, the monotonic behavior of $\f$ suggests that it may be interpreted as a generalized thermal $c$-function in the class of holographic systems considered here.

One important difference between the asymptotically AdS case and the flat case concerns the entropy bound~\eqref{Penrose}.  In the asymptotically AdS case, the Schwarzschild black hole has maximal horizon (entropy) only if we choose microcanonical ensemble (fixing total mass)~\cite{Marolf:2018ldl,Engelhardt:2019btp,Xiao:2022obq}. If we consider other ensembles, say, at fixed temperature, the situation may be different. For example, at zero temperature case planar Schwarzschild black brane has zero entropy, which is smaller than that of a charged black brane at the same temperature. For the static planar symmetric asymptotically AdS case, we argue that the Schwarzschild black brane may in fact have the smallest entropy,
\be
S(T) \ge S_{\rm Sch} (T) \, \label{S ineq-canonical0}
\ee
in the case of fixing temperature (canonical ensemble).

Before proceeding to our detailed results, let us emphasize the logical status from the outset. Our purpose in this paper is \textit{not to provide a general proof} of the local specific-heat bounds. Instead, we will use relatively simple tractable classes of solutions to identify and test a differential structure that unifies several known global inequalities. The local bounds are established below under explicit simplifying assumptions and are conjectural beyond that domain. By contrast, the integrated planar AdS inequality~\eqref{energy bound} admits a direct proof under the NEC for static asymptotically Schwarzschild-AdS black branes without assuming
$g_{tt}g_{rr}=-1$. The subsequent holographic consequences should therefore be read as consequences of this structural framework within its stated domain of validity.

The remainder of this paper is organized as follows. In Sec.~\ref{sec:asymptotic}, we focus on asymptotically flat black holes. We first provide evidence for the refined the specific heat bound \eqref{asy-golden-4D} under simplifying assumptions, and then demonstrate how it unifies various Penrose-like inequalities. In Sec.~\ref{sec:AdS}, we extend our framework to AdS backgrounds. We formulate the AdS specific heat bounds via a background subtraction procedure and support them with bulk local energy conditions. Subsequently, we explore the holographic implications of these bounds for the dual CFT and extract the macroscopic energy bound \eqref{energy bound}. Furthermore, we investigate the broad thermodynamic constraints imposed by this energy bound on holographic CFTs. We conclude the paper in Sec.~\ref{sec:conclusion}. We hope this work stimulates further attempts to prove or falsify these bounds in more general settings, rather than being viewed as a completed proof of all Penrose-like inequalities.

\section{Unification of geometric inequalities}
\label{sec:asymptotic}

In this section, we shall demonstrate that the Penrose-like inequalities associated with geometric thermodynamic quantities can arise from the refined specific heat bound \eqref{asy-golden-4D}. Before proceeding, we shall first try to establish the validity of \eqref{asy-golden-4D} in the next subsection by invoking an appropriate energy condition.

\subsection{A supporting derivation under simplifying assumptions}

The lower bound of \eqref{asy-golden-4D} was discussed in \cite{Lu:2025iky}; therefore, we shall focus on the upper bound. We begin with a general class of spherically-symmetric and static spacetimes in Einstein gravity minimally coupled to matter. The metric ansatz is given by
\be\label{generalg1}
\d s^2 = -e^{2\chi} f \d t^2 + f^{-1} \d r^2 + r^2 \d\Omega_2^2 \,,
\ee
where $f$ and $\chi$ are functions of the radial coordinate $r$, the ADM mass $M$, and the conserved charges $Q_i$. The event horizon is located at the largest root $r_h$ where $f(r_h)=0$. The temperature and entropy are $T = \fft{e^\chi f'}{4\pi} \big|_{r=r_h}$ and $S = \pi r_h^2$, respectively. Following the same procedure in \cite{Lu:2025iky}, it is straightforward to find that
\be
U = \fft{e^{\chi(r_h)}}{16\pi^2 r_h} \Big[ (2f'' + 2f'\chi') + rf'\pd_M \big(e^\chi f' \big) \Big]_{r=r_h} .
\ee
Substituting this geometric expression into the upper bound in \eqref{asy-golden-4D}, the thermodynamic inequality translates into a purely geometric constraint at the horizon
\be
e^{\chi(r_h)}\left[ (2f'' + 2f'\chi') + r f' \pd_M(e^\chi f') \right]_{r=r_h} \le \fft{4}{r_h^2} - \fft{6 e^{\chi(r_h)} f'(r_h)}{r_h} \,.
\label{asy-geo}
\ee

To prove \eqref{asy-geo}, we employ the Einstein equations $G_{\mu\nu}=8\pi T_{\mu\nu}$ with a diagonal matter energy-momentum tensor $T^\mu{}_\nu = \text{diag}\{-\rho, p_r, p_T, p_T\}$, where
\bea
\rho &=& \frac{1}{8\pi r^2} \big(1-f-rf' \big)\,,  \qquad  p_r = \frac{1}{8\pi r^2} \Big(rf'+f \big( 1 + 2 r \chi' \big) -1 \Big) \,, \nn \\
p_T &=& \frac{1}{16\pi r} \Big(f' \big(2 + 3r \chi'\big) + r f'' + 2 f \big(\chi' + r \chi'^2 + r \chi''\big) \Big)\,.\label{emtensor}
\eea
However, an analytical proof for the most general case ($\chi \neq 0$) is obstructed by the derivative term $\pd_M(e^\chi f')$. Local energy conditions provide algebraic relations among the metric functions and their spatial derivatives evaluated at a specific point, but they contain no information regarding derivatives with respect to the ADM mass $M$ unless the explicit matter Lagrangian is specified \cite{Lu:2025iky}. Consequently, we restrict our analytical proof to the class of static black holes where $g_{tt}g_{rr} = -1$ (i.e., $\chi = 0$). For such configurations, we have $\rho + p_r=0$; furthermore, the conservation of the matter energy-momentum tensor $\nabla_\mu T^\mu{}_\nu$ implies that
\be
\rho' + \frac{2}{r} \Big(\rho + p_T\Big)=0\,.
\ee
This consistent constraint on matter is independent of the metric profile function $f$, which, {\it a priori}, depends on the ADM mass $M$ of the solution. It is thus sensible to assume that the local energy density $\rho$ is independent of $M$. It follows from the first equation \eqref{emtensor} that
the function $f$ generally takes the form $f(r, M, Q_i) = 1 - 2M/r + \tg_{\text{flat}}(r, Q_i)$ \cite{Lu:2025iky}, where
\be
\tg_{\text{flat}}(r, Q_i) = \fft{1}{r} \int_{r}^\infty 8\pi \tilde r^2 \rho(Q_i,\tilde r)\, d\tilde r\,.
\ee
This immediately yields $\pd_M f'(r_h) = \fft{2}{r_h^2}$. The geometric inequality \eqref{asy-geo} then simplifies to
\be
\bigg(f'' + \fft{1}{r} f' \bigg)_{r=r_h} \le \fft{2}{r_h^2} - \frac{3}{r_h} f'(r_h) \,.
\ee
Using the Einstein field equations for $\chi=0$, it is straightforward to verify the identity
\be
\bigg(f'' + \fft{1}{r} f' \bigg)_{r=r_h} = \frac{2}{r_h^2} - \frac{3}{r_h} f'(r_h) - 16\pi(\rho - p_T)\Big|_{r=r_h} \,.
 \ee
Therefore, the refined upper bound holds provided that $\rho - p_T \ge 0$ at the horizon, which is precisely the DEC.

This derivation highlights the role of the temperature correction term $-3T/(2S)$. Without this term, the geometric inequality \eqref{asy-geo} would lack the $-3 f'(r_h)/r_h$ piece on the right-hand side. Saturation would then require both the saturation of the DEC ($\rho = p_T$) and the vanishing of the black hole temperature ($f'(r_h)=0$) \cite{Lu:2025iky}. By incorporating the correction, the explicit temperature dependence is absorbed, and the refined bound is saturated purely by the saturation of DEC at the horizon.

In addition, we note that while the analytical proof relies on $\chi=0$, this class encompasses the primary physical anchors for the Penrose-like inequalities, including the Schwarzschild and generic RN solutions. The original unrefined bound \eqref{asy-heatbound-4D} has been extensively confirmed in \cite{Lu:2025iky} across various $\chi \neq 0$ models using explicit black hole examples, and our refinement essentially optimizes the saturation behavior without altering the underlying thermodynamic structure. We therefore conjecture that the refined bound \eqref{asy-golden-4D} extends to general static black holes. As we shall see next, we can use the refined bound to derive the Penrose-like inequalities that were extensively tested by a variety of black holes.

\subsection{Unification of Penrose-like inequalities}

In the fixed-charge ensemble, thermodynamic variables such as mass $M$ and temperature $T$ are treated as functions of the entropy $S$. We now demonstrate how \eqref{asy-golden-4D} acts as a differential progenitor for the various Penrose-like inequalities through a systematic analysis of functional monotonicity. The logical structure of our derivation is summarized in Fig.~\ref{fig:roadmap-asy}. The differential inequalities below are evaluated along a fixed-$Q_i$ thermodynamic branch. To convert the resulting monotonicity statements into global algebraic inequalities, an additional endpoint/asymptotic condition is required. For the class of branches considered here, we assume that the appropriate endpoint is continuously connected to, or asymptotically approaches, the Schwarzschild relation. This endpoint assumption is logically distinct from the local bounds on $U$.
The monotonicity is then understood along this branch.

\begin{figure}[tbh]
	\centering
	\begin{tikzpicture}[node distance=2cm, auto, >=stealth]
		\node (L1) at (2.8, 1.5) {$U \ge -\dfrac{T}{2S}$};
		\node (L2) at (2.8, -0.5) {$U \le \dfrac{1}{4\sqrt{\pi S^3}} - \dfrac{3T}{2S}$};
		\node (A1) at (5.5, 3.0) {$X_1 = T^2S - \dfrac{1}{16\pi} \le 0$};
		\node (V1) at (8.7, 1.5) {$Y_1 = 3M - 2TS - \sqrt{S/\pi} \ge 0$};
		\node (V2) at (8.7, -0.5) {$Y_2 = M + 2TS - \sqrt{S/\pi} \ge 0$};
		\node (Final) at (13.0, 0.4) {$2M \ge \sqrt{S/\pi}$};
		\draw [->, thick] (2.8, 2.0) -- (3.8, 2.5);
		\draw [->, thick] (7.0, 2.5) -- (8.0, 2.0);
		\draw [->, thick] (3.8, 1.5) -- (6.0, 1.5);
		\draw [->, thick] (4.5, -0.5) -- (6.0, -0.5);
		\draw [thick, decorate, decoration={brace, amplitude=8pt}]
		(11.4, 1.6) -- (11.4, -0.9)
		node [midway, xshift=0.5cm] (brace_mid) {};
	\end{tikzpicture}
	\caption{\small The roadmap demonstrating how various Penrose-like inequalities are derived from the local specific heat bounds \eqref{asy-golden-4D} in asymptotically flat spacetimes. The arrows indicate the logical flow: the lower bound of $U$ ensures the monotonicity of the auxiliary function $X_1$, which subsequently yields the relaxed Penrose inequality $(Y_1 \ge 0)$. Similarly, the refined upper bound ensures the non-increasing behavior of $Y_2$, leading directly to the tightened Penrose inequality. Finally, the classical Penrose inequality is recovered by the algebraic addition of $Y_1$ and $Y_2$.}
	\label{fig:roadmap-asy}
\end{figure}

\subsubsection{Implications of the lower bound}

We first examine the consequences of the lower bound $U \ge -\fft{T}{2S}$. This constraint defines the monotonicity of an auxiliary function $ X_1 = T^2 S - \fft{1}{16\pi} $. By differentiating $X_1$ with respect to $S$ and utilizing the first law, we find
\be
\fft{\d X_1}{\d S} = 2TS \bigg( \fft{T}{2S} +U \bigg) \ge 0 \,.
\ee
Therefore, $X_1$ is a monotonically non-decreasing function of the entropy. For the fixed $Q_i$ branches considered here, we assume that the large-entropy regime asymptotically approaches the Schwarzschild thermodynamic relation, so that  $X_1\rightarrow0$ when $S\rightarrow\infty$. Consequently, the lower specific heat bound ensures $X_1 \le 0$ throughout the parameter space, which can also be equivalently written as
\be
2- \fft{1}{2T \sqrt{\pi S}} \le 0 \,.
\ee

For the relaxed Penrose inequality $Y_1 = 3M - 2TS - \sqrt{S/\pi} \ge 0$, it is easy to find that the derivative of $Y_1$ can be expressed as a linear combination of $X_1$ and the lower bound,
\be
\fft{\d Y_1}{\d S} = T \left( 2 - \fft{1}{2T \sqrt{\pi S}} \right) - 2S \left( \fft{T}{2S} + U \right) \le 0 \,.
\ee
Since $Y_1$ vanishes for the Schwarzschild solution, its global non-negativity is a direct consequence of this monotonic behavior. It is worth remarking on the connection between energy conditions and the strength of the resulting bounds. While the original geometric proof of $Y_1 \geq 0$ relies on the NEC \cite{Yang:2022yye}, our differential approach requires the strong energy condition (SEC) to ensure the lower bound of $U$ \cite{Lu:2025iky}. This more restrictive requirement is compensated by a stronger local statement: rather than mere global non-negativity, the SEC-driven bound ensures the monotonic evolution of $Y_1$ throughout the parameter space, which suggests that the relaxed Penrose inequality \eqref{Penrose-relax} is not a coincidental geometric property but is rooted in the local thermodynamic stability of the black hole.

\subsubsection{The refined upper bound}

Next we move to the refined upper bound in \eqref{asy-golden-4D}. Unlike the original loose upper bound in \eqref{asy-heatbound-4D}, the incorporation of the temperature correction allows for a direct derivation of the tightened Penrose inequality \eqref{Penrose-tighten}. By considering the auxiliary function $Y_2 = M + 2TS - \sqrt{S/\pi}$, we find that its derivative with respect to entropy is given by
\be
\fft{\d Y_2}{\d S} = 2S \left( U + \fft{3T}{2S} - \fft{1}{4\sqrt{\pi S^3}} \right).
\ee
The expression in the parentheses corresponds exactly to the refined upper bound. Thus, $\fft{\d Y_2}{\d S} \le 0$ identifies $Y_2$ as a monotonically non-increasing function. Under the same endpoint assumption,
$Y_2 \rightarrow0$ in limit $S\rightarrow\infty$. Thus, \eqref{Penrose-tighten} follows from the refined upper bound, once the above endpoint assumption is imposed.

A brief commentary on the underlying energy conditions provides further physical insight into this derivation. While the conventional geometric proof of $Y_2 \ge 0$ requires a combination of the NEC and the TEC \cite{Khodabakhshi:2022jot}, our differential framework relies exclusively on the DEC at the event horizon. The TEC restricts the trace of the energy-momentum tensor and is sensitive to the specific equation of state of the matter fields. In contrast, the DEC enforces causality by ensuring that the local energy flow is non-spacelike. The fact that the tightened Penrose inequality \eqref{Penrose-tighten} can be recovered via the DEC provides an alternative perspective, suggesting that this macroscopic bound is consistent with the causal structure of the matter fields.

Finally, as shown in Fig.~\ref{fig:roadmap-asy}, their algebraic sum is consistent with, and reproduces the form of, the classical Penrose inequality. This should not be interpreted as an independent proof of the classical Penrose inequality, because the derivation relies on the additional Schwarzschild endpoint/asymptotic assumption stated above. The differential framework established here demonstrates that the various algebraic bounds in black hole thermodynamics are not independent accidents but are manifestations of a single underlying specific heat bound \eqref{asy-golden-4D}. The utility of this inequality is further amplified when extended to AdS backgrounds, where it imposes thermodynamic constraints on the dual holographic CFT.

\section{Generalization to AdS black holes and holographic implications} \label{sec:AdS}

Having established the refined specific heat bound \eqref{asy-golden-4D} and its connection to geometric inequalities in asymptotically flat spacetimes, we now extend our framework to black holes in AdS backgrounds. This kind of generalization has its own physical significance. Operating through the AdS/CFT correspondence, we can translate these bulk bounds into explicit thermodynamic constraints on the dual CFT. As a result, these bounds, if they could be proved to be true in general, will offer a series of nontrivial constraints on the dual field theory. In the following discussion, we will use the boundary dual energy $E$ to replace the ADM mass $M$, since $E$ could be obtained from holographic renormalization naturally. For hyperbolic and planar symmetric cases, the energy $E$ and entropy $S$ will be infinite due to the infinite transverse volume. In this case we should talk about their densities. However, in order to maintain consistency and simplicity of formulas, we will still use $E$ and $S$, which should be understood as the finite densities multiplied by an infinite but fixed transverse volume.

\subsection{The AdS specific heat bounds}\label{sec31s}

\subsubsection{Generalizations in AdS case}

In the initial proposal \cite{Lu:2025iky} of the specific heat bound \eqref{asy-heatbound-4D}, the application to AdS spacetimes was treated in a heuristic manner. It was argued that because the negative cosmological constant characterizing the AdS vacuum violates the DEC, the upper bound of \eqref{asy-heatbound-4D} should be violated, leaving only the lower bound intact. However, this interpretation conflates the intrinsic properties of the background spacetime with those of the dynamical matter fields. In a well-defined physical setup, the ordinary matter fields coupled to gravity are expected to satisfy standard local energy conditions regardless of the asymptotic structure. The apparent violation of the DEC arises merely because the cosmological constant and the matter stress-energy tensor are lumped together into a single effective source in the standard Einstein equations.

To formulate a thermodynamic bound for AdS black holes, it is necessary to systematically subtract the contribution of the pure AdS vacuum, which isolates the thermodynamic contributions of the physical matter fields. Consequently, the saturation anchors of the inequalities are recalibrated to fit the AdS background in a manner consistent with the asymptotically flat case: the lower bound is saturated by the Schwarzschild-AdS solution, while the refined upper bound is simultaneously saturated by both the Schwarzschild-AdS and generic RN-AdS black holes. For a given bulk theory, such a subtraction could be realized by ``holographic renormalization''~\cite{Balasubramanian:1999re, deHaro:2000vlm,Bianchi:2001kw,Skenderis:2002wp}. Guided by these recalibrated saturation anchors, we finally propose these generalized specific heat bounds for four-dimensional static AdS black holes
\be
\fft{3}{4 \pi  \ell^2 \sqrt{\alpha  S}}-\fft{T}{2 S} \le U \le \fft{\sqrt{\alpha} k}{4 \pi S^{3/2}} -\fft{3T}{2S} +\fft{3}{2 \pi \ell^2 \sqrt{\alpha  S}}\,.
\label{ads-4D-k}
\ee
Here, the parameter $k=1, 0, -1$ characterizes the topology of the event horizon, corresponding to spherical, planar, and hyperbolic geometries, respectively, $\ell$ is the AdS radius, and entropy is assumed to be $S=\alpha r_h^2$, where $4\alpha$ is the spatial volume. As expected, the corrective terms introduced to the inequalities depend exclusively on the background scale $\ell$, remaining completely independent of the matter fields. Inequality~\eqref{ads-4D-k} recovers the flat case~\eqref{asy-golden-4D} in the limit of $\ell\rightarrow\infty$ with $k=1$.

To accommodate holographic dualities across various boundary dimensions, it is straightforward to extend the four-dimensional construction \eqref{ads-4D-k} to an arbitrary $(d+1)$-dimensional AdS bulk. Since the asymptotically flat bound \eqref{asy-heatbound-4D} has already been generalized to higher dimensions \cite{Lu:2025iky}, we can apply a similar background subtraction method. By anchoring the saturation conditions to the $(d+1)$-dimensional Schwarzschild-AdS and RN-AdS solutions, the generalized AdS bounds read
\be
\fft{d \alpha^{\fft{1}{1-d}} S^{\fft{2-d}{d-1}}}{2 \pi  (d-1) \ell^2} -\fft{1}{d-1} \fft{T}{S} \le U \le \fft{(d-2)^2 k \alpha ^{\fft{1}{d-1}} S^{\fft{d}{1-d}}}{2 \pi  (d-1)} -\fft{(2d-3)T}{(d-1)S} +\fft{d \alpha ^{\fft{1}{1-d}} S^{\fft{2-d}{d-1}}}{2 \pi \ell^2}\,,
\label{ads-D-k}
\ee
where $d$ denotes the spacetime dimension of the dual boundary field theory. Although these generalized inequalities apply to all horizon topologies, the planar case with $k=0$ is of particular conceptual and practical significance. Within the AdS/CFT dictionary, planar black holes are dual to thermal states in a CFT residing in a flat spacetime. In the subsequent analysis, we shall restrict our attention to the planar solutions.

Before proceeding to the analytical proof, it is illuminating to examine the lower bound of \eqref{ads-D-k} in the planar limit. The first term on the left-hand side is purely geometric, depending explicitly on the AdS radius $\ell$ and the spatial volume parameter $\alpha$. The second term, conversely, is purely thermodynamic. For the Schwarzschild-AdS black brane, which is characterized by the absence of generic matter fields and saturates this lower bound, we find the following algebraic relation
\be
U_\text{Sch} = \fft{d \alpha^{\fft{1}{1-d}} S^{\fft{2-d}{d-1}}}{4 \pi  (d-1) \ell^2} = \fft{T}{(d-1)S} \,.
\ee
This relation suggests that we can construct two distinct lower bounds for the quantity $U$, both of which are saturated by the Schwarzschild-AdS black brane. We define these as the geometric and thermodynamic lower bounds respectively
\be
U_{\rm geo} = \fft{d \alpha^{\fft{1}{1-d}} S^{\fft{2-d}{d-1}}}{4\pi(d-1)\ell^2} \,, \qquad  U_{\rm the} = \fft{1}{d-1}\fft{T}{S} \,.
\label{two lower bounds}
\ee
The proposed constraints then take the form $U \ge U_{\rm geo}$ and $U \ge U_{\rm the}$. While these two bounds coincide for the Schwarzschild-AdS black brane, they diverge in the presence of additional matter fields. For the RN black brane, it can be explicitly verified that $U_{\rm the} \le U_{\rm geo}$, which implies that the geometric bound provides a more stringent constraint than the purely thermodynamic one.

In the following subsection, we employ local energy conditions in some special cases to support the following thermodynamic hierarchy (here we focus on planar case $k=0$)
\be
U_{\rm the} \le U_{\rm geo} \le U \le U_{\text{upper}},~~U_{\text{upper}}=-\fft{(2d-3)T}{(d-1)S} +\fft{d \alpha ^{\fft{1}{1-d}} S^{\fft{2-d}{d-1}}}{2 \pi \ell^2}\,,
\label{ads-D-planar}
\ee
which reveals a coherent interplay between bulk energy conditions and the constraints on $U$. Without additional statement, we will also assume that the spacetime is asymptotically Schwarzschild-AdS rather than only asymptotically AdS. This condition requires the matter sector in the energy-momentum tensor decays faster than $\mathcal{O}(1/r^d)$ near the boundary.\footnote{See Ref.~\cite{Yuguang} for a definition about what is ``asymptotically Schwarzschild-AdS''.} In such a setup, the ``holographic renormalization'' reduces into a simple background subtraction by using vacuum AdS spacetime.  Though it is not general enough to cover all physically interesting cases, it simplifies the complication caused by holographic renormalization and allows us to focus on our inequality itself.

\subsubsection{Support inequalities with bulk energy conditions}
The procedure to establish \eqref{ads-D-planar} via local energy conditions closely parallels the derivation in asymptotically flat spacetimes. As demonstrated in Sec.~\ref{sec:asymptotic}, evaluating $U$ for a fully general metric introduces derivatives with respect to the mass parameter $M$ that cannot be constrained by local stress-energy tensors. To avoid redundant exposition, we bypass the generic obstruction associated with $g_{tt}g_{rr} \neq -1$ and directly consider the class of static, planar-symmetric black holes satisfying $g_{tt}g_{rr} = -1$ in this subsection. Though such an assumption means that our following discussion falls short of rigorous proof, it simplifies a large number of technical details and offers a useful evidence.

The corresponding $(d+1)$-dimensional metric ansatz is given by
\be
\d s^2= -r^2 f(r) \d t^2 + \fft{\d r^2}{r^2 f(r)} + r^2 \d \vec{x}^2_{d-1}\,, \qquad f= 1/\ell^2 -f_0/r^d +\tg_{\text{AdS}}(r, Q_i)\,,
\label{metric-1}
\ee
where $\d\vec{x}_{d-1}^2$ denotes the line element of the $(d-1)$-dimensional flat transverse space, and $Q_i$ represent possible source or charge parameters. The parameter $f_0$ represents the integration constant associated with the energy density of the black brane, playing a role analogous to the mass term $2M$ in the asymptotically flat case. The outer event horizon is located at $r=r_h$, defined by the largest root of $f(r_h)=0$. For this geometry, the energy, temperature and entropy are given by
\be
E = \fft{(d-1)\alpha}{4\pi} f_0 \,, \qquad  T = \fft{r_h^2 f'(r_h)}{4\pi} \,, \qquad S = \alpha r_h^{d-1} \,, \label{thermo-quantities}
\ee
where $4\alpha$ represents the spatial volume factor of the transverse space. Moreover, with the metric ansatz \eqref{metric-1}, the matter energy-momentum tensor is diagonal and takes the form $T^\mu_\nu = \text{diag}\{-\rho(r), p_r(r), p_T(r), \cdots, p_T(r)\}$. The Einstein field equations reduce to
\be
\rho =-p_r =-\fft{d-1}{16\pi} \big(d {} f +rf' \big)\,, \quad p_T= \fft{1}{16\pi} \big[(d-1)d {} f +2d {} r {} f' +r^2 f'' \big]\,.
\ee
The total energy-momentum can be decoupled into a matter term and a background term, e.g., $\rho = \rho_m + \rho_\Lambda$. Substituting $f$ into the energy momentum tensor, we find that the background terms satisfy the following relation
\be
\rho_\Lambda = -p_{r_\Lambda} = -p_{T_\Lambda} = -\frac{(d-1) d}{16 \pi  \ell^2} \,.
\ee

Instead of proving the entire hierarchy simultaneously, we tackle the sequence of inequalities in \eqref{ads-D-planar} step by step, which reveals a direct correspondence with the standard energy conditions. We begin with the thermodynamic lower bound $U_{\rm the}$. By computing the difference between $U$ and $U_{\rm the}$ at the horizon, we find
\be
U-U_{\rm the} = \fft{4r_h^{2-d}}{(d-1)\alpha} \big( \rho+p_T \big) \Big|_{r=r_h} = \fft{4r_h^{2-d}}{(d-1)\alpha} \big( \rho_m +p_{T_m} \big) \Big|_{r=r_h} \,.
\ee
This demonstrates that the thermodynamic lower bound is protected by the NEC ($\rho_m +p_{T_m} \ge 0$) of the dynamical matter fields at the event horizon. In addition, recalling the definition of $U$ \eqref{U} on the positive-temperature branch for which $U>0$, this can equivalently be expressed as,
\be
C_{Q_i} \le (d-1)S \,. \label{ads-CQ-S}
\ee

Next, we evaluate the geometric lower bound $U_{\rm geo}$. The difference between $U$ and $U_{\rm geo}$ is governed by a different combination of the stress-energy components
\bea
U-U_{\rm geo} &=& \fft{4r_h^{2-d}}{(d-1)^2 \alpha} \big[ (d-2)\rho +(d-1)p_T -p_{T_\Lambda} \big] \Big|_{r=r_h} \nn \\
&=& \fft{4r_h^{2-d}}{(d-1)^2 \alpha} \big[ (d-2)\rho_m +(d-1)p_{T_m} \big] \Big|_{r=r_h} .
\eea
This implies that the geometric lower bound is enforced when the matter sector satisfies both the SEC ($(d-2)\rho_m +\sum_i p_{i_m} \ge 0$, where $i$ denotes all spatial components) and the weak energy condition (WEC, $\rho_m \ge 0$) at the horizon. The structural relationship between these two lower bounds can be clarified by directly comparing $U_{\rm the}$ and $U_{\rm geo}$. The subtraction of these two quantities isolates the matter energy density
\be
U_{\rm geo} - U_{\rm the} = \fft{4r_h^{2-d}}{(d-1)^2 \alpha} \big( \rho-\rho_\Lambda \big) \Big|_{r=r_h}
= \fft{4r_h^{2-d}}{(d-1)^2 \alpha} \rho_m \Big|_{r=r_h}  \,.
\ee
Thus, the geometric bound dominates the thermodynamic one provided the matter energy density is non-negative, corresponding precisely to the WEC. It is worth noting that the satisfaction of the WEC and SEC naturally necessitates the NEC, implying that the thermodynamic bound $U_{\rm the}$ is inherently more fundamental than the geometric bound $U_{\rm geo}$.

Finally, we examine the upper bound in \eqref{ads-D-planar}. It is straightforward to establish that
\bea
U -U_{\text{upper}} &=& \fft{4r_h^{2-d}}{(d-1) \alpha}
\big[ (p_T -\rho) +2\rho_\Lambda] \Big|_{r=r_h} \nn \\
&=& \fft{4r_h^{2-d}}{(d-1) \alpha} \big( p_{T_m} -\rho_m \big) \Big|_{r=r_h} \,,
\eea
which ensures that the upper bound holds true whenever the DEC ($\rho_m -p_{T_m} \ge 0$) is satisfied at the horizon. Because the DEC implies the NEC, any physical configuration satisfying the upper bound must automatically respect the lower bound $U \ge U_{\rm the}$. This confirms that $U_{\rm the}$ acts as the minimal thermodynamic requirement for these planar AdS black holes.

To conclude, the step-by-step derivation presented above uncovers the hierarchical structure Eq.~\eqref{ads-D-planar} that directly reflects the strength of the local energy conditions, which can be summarized as

\be
\xymatrix@C=5em{
U_{\rm upper}&\ar[l]^{~~\geq~}_{~~{\rm DEC}}U&\ar[l]^{~~\geq}_{~{\rm SEC+WEC}}U_{\rm geo}&U_{\rm the}\ar[l]^{~~\geq}_{~{\rm WEC}} \ar`d/0pt[d][lld]_{{\rm NEC}}^{\geq}\\
& \ar[u]& &
}
\label{hierarchy U}
\ee
In this way, the AdS specific heat bounds translate the standard bulk energy conditions into a nested family of local thermodynamic inequalities. Interestingly, the $U_{\rm geo}\geq U_{\rm the}$ requires WEC and $U\geq U_{\rm geo}$ requires ``WEC+SEC'', but their ``corollary''  $U\geq U_{\rm the}$ only needs the weaker energy condition NEC. It should be noted that all derivations in this subsection are given under the assumption $g_{rr}g_{tt}=-1$, so the rigorous proofs are still absent for the bounds on U in Eq.~\eqref{ads-D-planar}. However, our above analyses offer strong nontrivial evidence and we hope we can obtain more general proofs for inequalities~\eqref{ads-D-planar} in the future. In the following subsection, we will demonstrate that these local bulk bounds translate directly into macroscopic thermodynamic constraints on the dual CFT through the AdS/CFT correspondence.

\subsection{Holographic implications}

The analytical derivation in the previous subsection demonstrates that the thermal properties of bulk black holes can be related to bulk energy conditions under the assumptions specified above. We now explore the global consequences of these bulk constraints on the thermodynamic structure of holographic CFTs. Specifically, we will demonstrate how the bulk bounds enforce a macroscopic energy bound \eqref{energy bound} and restrict the normalized entropy density. There are several different ways to derive the bound \eqref{energy bound}.

Perhaps the simplest derivation starts from inequality~\eqref{ads-CQ-S}. To see that, we define $\Theta(T)=(E-E_0)d-(d-1)TS$. The derivative with respect to $T$ gives us $\Theta'= E'd-(d-1)(S+TS')$ and so we find
\begin{equation}\label{verifyeqts1}
  \Theta'=TS'd -(d-1)(S+TS')=C_{Q_i}-(d-1)S\,.
\end{equation}
Here we have used fixed-charge first law $E'=TS'$ and the fact $E_0$ is constant. Thus, inequality~\eqref{ads-CQ-S} implies $\Theta'\leq0$. On the other hand, we note that the bulk solution approaches the Schwarzschild-AdS black brane in the high-temperature limit, and so we have $\lim_{T\rightarrow\infty}\Theta(T)=0$. In the high-temperature regime, the scales associated with the deformation or fixed charges become parametrically small compared with $T$. Under the assumed asymptotically Schwarzschild-AdS boundary conditions, the solution therefore approaches the corresponding Schwarzschild-AdS black brane, so that $\Theta(T)\rightarrow0$. Thus, we should expect $\Theta\geq0$ for finite $T$ and we then obtain bound \eqref{energy bound}. Of course, this is not a rigorous proof for bound \eqref{energy bound}, since the inequality~\eqref{ads-CQ-S} has not been proven, and the approach to the Schwarzschild-AdS black brane at high temperature limit is a ``physical picture'' not a rigorous ``theorem''.

Since it has been conjectured that NEC could guarantee the inequality~\eqref{ads-CQ-S}, our above argument in fact predicts a pure geometry inequality that could be tested in classical gravity theory without referring to the AdS/CFT correspondence. For static planar asymptotically Schwarzschild-AdS black branes in Einstein gravity, the bulk NEC is sufficient to establish~\eqref{energy bound}, as proven independently in Appendix~\ref{adsets1}. This independent proof of~\eqref{energy bound} in Appendix~\ref{adsets1} provides a nontrivial consistency check for the broader differential picture suggested by~\eqref{ads-CQ-S}, although it does not by itself establish the local bound~\eqref{ads-CQ-S}.

In the following, we discuss some physical implications of bound~\eqref{energy bound}  in holography.  We further show that \eqref{energy bound} leads to the monotonic behavior of the normalized free energy and imposes restrictions on both the canonical free energy and the microcanonical entropy.

\subsubsection{Bounds on normalized entropy density}

Having extracted the macroscopic energy constraint \eqref{energy bound} from the thermodynamic lower bound \eqref{ads-CQ-S}, we first discuss the implications of the full specific heat bounds \eqref{ads-D-planar}. Specifically, we will demonstrate how these local bounds restrict the normalized entropy density $\tc(T)$ of the holographic CFT.

In a $d$-dimensional holographic CFT, the central charge $c$ is related to the bulk gravitational parameters \cite{Liu:1998bu,Kovtun:2008kw}, which is given by (in the unit $\ell=G_N=1$)
\be
c = \fft{d+1}{d-1} \fft{\Gamma(d+1)\pi^{d/2}}{4\pi \Gamma(d/2)^3} \,.
\ee
Inspired by pure CFTs, we define the normalized entropy density $\tc(T)$ \cite{Kovtun:2008kw,Kovtun:2008kx} by factoring out the explicit temperature dependence and a specific geometric constant $s_d$
\be
\tc(T) = \fft{S}{4\alpha s_d T^{d-1}\ell^{2(d-1)}} \,, \quad \text{where} \quad s_d \equiv \fft{\Gamma(d/2)^3}{4\pi^{d/2}\Gamma(d)} \left( \fft{4\pi}{d} \right)^d \fft{d-1}{d+1} \,.  \label{def.normalized entropy}
\ee
The constant $s_d$ is carefully chosen such that in the exact conformal limit, $\tc$ precisely recovers the central charge $c$. Here we keep the notation $\ell$ for convenience. This conformal limit corresponds to the ultraviolet (UV) regime, which is dominated by the Schwarzschild-AdS black brane, yielding $\lim_{T \to \infty} \tc(T) = c$. For deformed systems at finite temperature, $\tc(T)$ exhibits non-trivial temperature dependence. Taking the temperature derivative of $\tc(T)$ in the fixed-charge ensemble, we find
\be
\left( \fft{\pd \tc}{\pd T} \right)_{Q_i} = \fft{1}{4\alpha s_d T^d\ell^{2(d-1)}} \Big( C_{Q_i} - (d-1)S \Big) \,.
\ee
Invoking the thermodynamic lower bound \eqref{ads-CQ-S}, we immediately obtain
\be
\left( \fft{\pd \tc}{\pd T} \right)_{Q_i} \le 0 \,.
\ee
Thus, the normalized entropy density decreases monotonically with temperature. While $\tc(T)$ acts as a monotonic function, the non-positive sign of its derivative indicates that it behaves in the opposite direction to a traditional holographic $c$-function along the RG flow \cite{Zamolodchikov:1986gt}. Combined with the fact that  $\tilde{c}(\infty)=c$, this gives
\be
\tc(T) \ge c \,, \label{tc-c}
\ee
which demonstrates that any additional matter fields will necessarily increase the entropy density of the system relative to the pure gravitational vacuum at the same temperature.
Therefore, we immediately obtain
\be
S(T) \ge S_{\rm Sch} (T) \,. \label{S ineq-canonical}
\ee
The equality is saturated exclusively by the Schwarzschild-AdS black brane, which means that the right-hand side of this inequality is exactly the entropy of a pure Schwarzschild-AdS state at a given temperature $T$.

At first glance, this inequality contrasts with the original Penrose inequality \eqref{Penrose}, which establishes the Schwarzschild solution as the state of maximum entropy. The distinction arises purely from the choice of thermodynamic ensemble. The original Penrose inequality is a microcanonical statement operating at fixed energy, whereas the bound here applies to the canonical ensemble at fixed temperature and the charge $Q_i$.

We can also obtain inequality~\eqref{S ineq-canonical} in a different way, which provides a comprehensive view of \eqref{ads-D-planar}. Let us consider the refined geometric upper bound $U \le U_{\text{upper}}$. Following a similar differential approach, we construct another auxiliary function $\tilde{g}(T)$ tailored for arbitrary $(d+1)$ dimensions
\be
\tilde{g}(T) = \fft{d\alpha^{\fft{1}{1-d}}}{4\pi\ell^2} S^2 - T S^{\fft{2d-3}{d-1}} \,.
\ee
Taking the temperature derivative in the fixed-charge ensemble yields
\be
\tilde{g}'(T) = - S^{\fft{2d-3}{d-1}} U^{-1} \left( U - U_{\rm upper} \right) \,,
\ee
where $U_{\rm upper}$ denotes the refined upper bound in \eqref{ads-D-planar}. The inequality $U \leq U_{\rm upper}$, combined with the thermodynamic lower bound $U\geq U_{\text{the}}\geq0$, ensures $\tilde{g}'(T) \geq 0$. Therefore, $\tilde{g}(T)$ is a monotonically non-decreasing function of temperature along fixed-charge branches. Bounded from below by its zero-temperature limit $\tilde{g}(0) \ge 0$, the non-negativity of $\tilde{g}(T)$ imposes an algebraic restriction on the entropy~\eqref{S ineq-canonical}.

The logical flow of the thermodynamic consequences derived thus far is summarized in Fig.~\ref{fig:roadmap-ads1}. As the diagram shows, there is a structural convergence between the upper and lower bounds. The thermodynamic lower bound \eqref{ads-CQ-S} and the  refined upper bound $U \leq U_{\rm upper}$ provide two independent analytical pathways that culminate in the exact same entropy constraint \eqref{S ineq-canonical}. Having established how the local specific heat bounds restrict the entropy density, our focus now shifts to the macroscopic energy bound \eqref{energy bound} to explore the deeper thermodynamic structures it governs.

\begin{figure}[h]
	\centering
	\begin{tikzpicture}[node distance=2cm, auto, >=stealth]	
		\node (lower) at (-1.3, 1.75) { $\mathrm{NEC}\Rightarrow C_{Q_i} \le (d-1)S \Rightarrow$};
		\node (upper) at (-0.5, -0.8) { $\mathrm{DEC}\Rightarrow U \le U_{\text{upper}}$};
		\node (energy) at (3.5, 2.5) { $d(E - E_0) \ge (d-1)TS$};
		\node (monotonicity) at (3.5, 1.0) { $\dfrac{\pd \tc}{\pd T} \le 0$};
		\node (entropy-bound) at (3.6, -0.8) { $S(T) \ge S_{\text{Sch}}(T)$};
		\draw[thick, decorate, decoration={brace, amplitude=8pt}] (1.3, 0.9) -- (1.3, 2.6) node[black, midway, xshift=0.6cm] {};
		\draw [->, thick] (1.1, -0.8) -- (1.8, -0.8);
		\draw [->, thick] (3.7, 0.6) -- (3.7, -0.5);
		
	\end{tikzpicture}
	\caption{The roadmap of the holographic thermodynamic constraints derived from \eqref{ads-D-planar}.}
	\label{fig:roadmap-ads1}
\end{figure}

\subsubsection{Bounds on free energy density}
The quantity $\tc$ characterizes the entropy of the system, the characteristic thermodynamic quantity of an isolated system, usually used in the microcanonical ensemble. If we turn to a system at fixed temperature but able to exchange energy with an ``environment'', we have to use the canonical ensemble. In this case, the Helmholtz free energy $F$ replaces entropy as the stability criterion. The standard thermodynamic relations dictate that
\be
E = F + TS\,, \qquad S = -\pd F / \pd T \,. \label{thermo-relation EF&SF}
\ee
Let us consider how our previous inequalities constrain the behaviors of the free energy.

Guided by the scaling behavior of pure CFTs, we factor out the exact geometric constants and express the free energy in terms of a dimensionless, normalized free energy density $\f(T)$ defined by the relation
\be
F(T) = E_0 - \fft{4s_d \alpha \ell^{2(d-1)}T^d}{d} \f(T) \,,
\ee
where $E_0$ is the reference energy of the vacuum. Here we focus only on planar symmetric static asymptotically Schwarzschild-AdS spacetime, so we can set $E_0=0$. However, we still keep the notation $E_0$ in formulas. By applying the thermodynamic relations \eqref{thermo-relation EF&SF} to this definition, it is straightforward to establish an algebraic identity connecting the macroscopic energy bound \eqref{energy bound} directly to the temperature derivative of $\f(T)$
\be
d(E - E_0) - (d-1)TS = \fft{4\alpha s_d \ell^{2(d-1)}T^{d+1}}{d} \fft{\pd \f}{\pd T} \label{tildefe1}\,.
\ee
The positivity of the left-hand side is guaranteed by the lower specific heat bound \eqref{ads-CQ-S} derived from the bulk NEC. Consequently, we immediately obtain $\pd \f/\pd T \ge 0$, establishing that $\f$ is a monotonically non-decreasing function of temperature. This monotonic behavior complements the decreasing nature of the normalized entropy density $\tc(T)$. Utilizing the definition of $\tc$ \eqref{def.normalized entropy}, one can easily verify the relation
\be
\tc = \f + \fft{T}{d} \fft{\pd \f}{\pd T} \,,
\ee
which immediately ensures $\tc(T) \ge \f(T)$.

To uncover the physical significance of $\f(T)$, we assume that the boundary field theory possesses both a UV fixed point and an infrared (IR) fixed point. In order to admit a valid gravity dual, it has been argued in \cite{Kovtun:2008kw} that $\tc(T)$ must exactly match the corresponding central charges at these conformal fixed points, namely $\tc(\infty) = c_{\rm UV}$ and $\tc(0) = c_{\rm IR}$. Furthermore, conformal symmetry near these fixed points implies that the thermal free energy scales as $F - E_0 \propto T^d$. This scaling implies that $\f(T)$ must approach a constant, forcing the derivative term $\pd \f/\pd T$ to vanish at the endpoints. Consequently, we find $\f(\infty) = c_{\rm UV}$ and $\f(0) = c_{\rm IR}$. The monotonicity $\pd\f/\pd T \ge 0$ then enforces the hierarchy $c_{\rm IR} \le c_{\rm UV}$. This suggests that $\f$ acts as a thermal analogue of Zamolodchikov's $c$-function \cite{Zamolodchikov:1986gt}, which mimics the monotonic reduction of effective thermal degrees of freedom along the RG flow.

In addition, the monotonic growth of $\f(T)$ towards its UV limit $c_{\rm UV}=c$ algebraically implies that $\f(T) \le c$ everywhere in the parameter space, which can be translated to the global inequality
\be
F(T) \ge F_{\text{Sch}}(T) \,, \label{F ineq-canonical}
\ee
where $F_{\text{Sch}}$ denotes the free energy of the pure Schwarzschild-AdS black brane. It is instructive to consider this result alongside our earlier observation \eqref{S ineq-canonical}. Together, they reveal that the Schwarzschild-AdS black brane simultaneously minimizes both the free energy and the entropy among all permissible configurations in the canonical ensemble.

This thermodynamic structure extends to the microcanonical ensemble through the ratio $S^d / (E-E_0)^{d-1}$, whose partial derivative with respect to $S$ is given by
\be
\fft{\pd}{\pd S} \left( \fft{S^d}{(E-E_0)^{d-1}} \right) = \fft{S^{d-1}}{(E-E_0)^d} \Big[ d (E-E_0) - (d-1)TS \Big] \,.
\ee
The macroscopic energy bound \eqref{energy bound} ensures that this derivative is non-negative, identifying the ratio as a monotonically non-decreasing function of the entropy. In the limit $S \to \infty$, the temperature also diverges because the normalized entropy density \eqref{def.normalized entropy} remains finite. In this UV regime, the system asymptotically recovers the Schwarzschild-AdS black brane geometry, where the ratio approaches a fixed geometric constant. Monotonicity therefore requires the ratio for any finite state to be bounded by its value in the conformal limit
\be
\fft{S(E)^d}{(E-E_0)^{d-1}} \le \lim_{S \to \infty} \fft{S(E)^d}{(E-E_0)^{d-1}} = \fft{S_\text{Sch}(E)^d}{E_\text{Sch}^{d-1}}\,.
\ee
Here the $E_0$ disappears from the rightmost expression since $E_0=0$. This remains true for the case $E_0\neq0$ since $E\rightarrow\infty$ under the limit $S\rightarrow\infty$. By matching the energy of a Schwarzschild-AdS black brane to the generic state such that $E_{\text{Sch}} = E$, we obtain $S(E) \le S_{\text{Sch}}(E)$. This identifies the Schwarzschild-AdS black brane as the state of maximum entropy for a given energy, providing a purely thermodynamic derivation of the classical Penrose inequality \eqref{Penrose} in the AdS backgrounds.

The thermodynamic inequalities derived in this subsection are summarized in Fig.~\ref{fig:roadmap-ads2}. This diagram captures the systematic propagation of the thermodynamic lower bound \eqref{ads-CQ-S} into the global properties of the dual CFT. The resulting macroscopic energy bound \eqref{energy bound} acts as a central node branching into distinct thermodynamic ensembles. In the canonical ensemble, it enforces the monotonicity of $\f$, which subsequently leads to the bounds on $\tc$ and the free energy itself. Simultaneously, in the microcanonical ensemble, it provides the foundation for recovering the classical Penrose inequality.

\begin{figure}[h]
	\centering
	\begin{tikzpicture}[node distance=2cm, auto]
		\node (U) at (-1.1, 0.5) { $\mathrm{NEC}\Rightarrow C_{Q_i} \le (d-1)S $};
		\node (E) at (4, 0.5) { $d(E - E_0) \ge (d-1)TS\Rightarrow$};
		
		\node (df) at (7.5, 1.5) { $\dfrac{\pd \f}{\pd T} \ge 0$};
		\node (S) at (8.7, -0.5) { $S(E) \le S_{\text{Sch}}(E)$};
		
		\node (c) at (9.5, 2.2) { $\tc \ge \f$};
		\node (f) at (10.5, 0.8) { $F(T) \ge F_{\text{Sch}}(T)$};
		\draw [->, thick] (1.0, 0.5) -- (1.7, 0.5);
		\draw[thick, decorate, decoration={brace, amplitude=8pt}] (6.5, -0.5) -- (6.5, 1.5) node [black, midway, xshift=0.5cm] {};
		\draw [thick, decorate, decoration={brace, amplitude=6pt}] (8.7, 0.8) -- (8.7, 2.2) node[black, midway, xshift=0.5cm] {};
		
	\end{tikzpicture}
	\caption{The roadmap illustrating the holographic constraints derived from \eqref{energy bound}.}
	\label{fig:roadmap-ads2}
\end{figure}

We conclude this subsection by emphasizing the underlying significance of these derived relationships. While an arbitrary boundary field theory may possess its own thermodynamic structure, it is not expected to obey these particular constraints. However, for a strongly coupled CFT admitting a semiclassical gravitational dual, the bulk energy conditions translate into nontrivial thermodynamic constraints on the boundary theory. From this perspective, the thermodynamic inequalities derived here should be viewed as constraints closely tied to the existence of a semiclassical gravitational dual. The holographic consequences discussed above follow algebraically once the relevant thermodynamic bounds are assumed. The remaining open issue concerns primarily the validity of the local specific-heat bounds beyond the restricted classes of bulk solutions considered in Sec.~\ref{sec31s}. Establishing or falsifying these local bounds in more general settings would be an interesting direction for future work.

\section{Conclusion} \label{sec:conclusion}
In this paper, we identify a local-to-global thermodynamic mechanism that organizes several previously disconnected inequalities. We establish it analytically in controlled settings, obtain one stronger integrated result under weaker assumptions, and formulate the general local statements as concrete open problems. For asymptotically flat spacetimes, we proposed a refined upper bound \eqref{asy-golden-4D} on the inverse specific heat $U$. By explicitly evaluating the local energy conditions on the event horizon, we demonstrated that \eqref{asy-golden-4D} guarantees the functional monotonicity required to systematically unify various Penrose-like inequalities. This differential framework suggests that macroscopic geometric bounds, such as the relaxed \eqref{Penrose-relax} and tightened Penrose inequalities \eqref{Penrose-tighten}, are not isolated geometric coincidences but instead emerge from the causal and energy structures of the underlying matter fields.

We subsequently extended this thermodynamic framework to AdS backgrounds. To isolate the thermodynamic contributions of the physical matter fields from the negative cosmological constant, we implemented a background subtraction procedure. This approach yielded generalized specific heat bounds for planar AdS black holes \eqref{ads-D-planar} that organize into a hierarchy \eqref{hierarchy U} controlled by the null, weak, strong, and dominant energy conditions. Particularly, we found that inequality~\eqref{ads-CQ-S} may be ensured by NEC. We then discussed some applications of inequality~\eqref{ads-CQ-S} to the AdS/CFT correspondence.  Through the AdS/CFT correspondence, we translated these local bulk constraints into global thermodynamic restrictions on the dual CFT, culminating in a macroscopic energy bound \eqref{energy bound}. These restrictions are summarized in Fig.~\ref{fig:roadmap-ads1}. A central consequence of \eqref{energy bound} is that it imposes nontrivial thermodynamic constraints on the holographic CFT, as summarized in Fig.~\ref{fig:roadmap-ads2}. By operating in the canonical ensemble, combining the NEC-based integrated bound~\eqref{energy bound} with the thermodynamic identity~\eqref{tildefe1}, we find that the bulk NEC enforces the monotonic evolution of a normalized free energy density $\f(T)$ with respect to temperature, which suggests that $\f(T)$ behaves as a generalized thermal $c$-function within the holographic framework considered here. We use the term ``thermal $c$-function'' only in the limited sense of endpoint matching and monotonicity with temperature; no identification of temperature evolution with a genuine RG trajectory is assumed.

The status of the various bounds is not identical. The local bounds on $U$ are supported analytically under simplifying assumptions such as
$g_{tt}g_{rr}=-1$ and matter profiles independent of the ADM mass, whereas the integrated AdS bound~\eqref{energy bound} admits a direct proof for static planar asymptotically Schwarzschild-AdS black branes satisfying the NEC, without imposing either of the above two assumptions.  Within these explicit assumptions our derivations are self-consistent, and the saturation by Schwarzschild and Reissner-Nordstr\"{o}m families, together with the consistency checks summarized in Figs. \ref{fig:roadmap-asy}-\ref{fig:roadmap-ads2}, provide nontrivial evidence that the underlying differential structure may persist beyond this restricted setting.

A fully general proof, particularly for $\chi\neq0$, rotating, and higher-derivative configurations, remains an open problem, which we leave for future study. Beyond the individual inequalities, the main conceptual outcome of our analysis is the common local-to-global structure summarized in Figs.~\ref{fig:roadmap-asy}-\ref{fig:roadmap-ads2}. This structure suggests that global geometric inequalities may be viewed as integrated manifestations of local thermodynamic constraints. Traditionally, macroscopic algebraic relations in black hole thermodynamics, such as the classical Penrose inequality \eqref{Penrose}, have been established through geometric analyses of initial data sets \cite{Penrose:1973um,Bray:2003ns,Mars:2009cj,Ben-Dov:2004lmn,Bray:1999eow,Huisken:2001,Hayward:1994bu,Visser:1992qh,Yang:2020ark,Dain:2017jkj}. The findings in this paper provide a complementary perspective: these global algebraic relations can emerge as the integrated consequences of local differential bounds on the specific heat. Looking forward, it would be instructive to investigate whether these specific heat bounds survive the inclusion of higher-derivative corrections \cite{Lu:2015cqa,Lu:2015psa} or semiclassical backreactions \cite{Wall:2011hj,Engelhardt:2014gca}. Furthermore, extending this differential framework to stationary rotating black holes would also provide a more comprehensive understanding of the intricate interplay between geometric inequalities and black hole thermodynamics.

\section*{Acknowledgement}

This work is supported in part by the National Natural Science Foundation of China (NSFC) grants No.~12375051, No.~12375052 and No.~11935009, and also by the Tianjin University Self-Innovation Fund Extreme Basic Research Project Grant No.~2025XJ21-0007 and No.~2025XJ22-0014.

\appendix

\section{Proof of \eqref{energy bound} for static planar-symmetric asymptotically Schwarzschild-AdS black branes}\label{adsets1}
In this appendix, we will supply a proof of the bound~\eqref{energy bound} for planar-symmetric asymptotically Schwarzschild-AdS black branes. Particularly, we here do not assume $g_{tt}g_{rr}=-1$. We consider the metric
\be
\d s^2= -r^2 f(r)e^{2\chi} \d t^2 + \fft{\d r^2}{r^2 f(r)} + r^2 \d \vec{x}^2_{d-1}\,,
\label{metric-ads2}
\ee
We take AdS radius $\ell=1$. For the black hole case, there is an event horizon at $r=r_h$. Here $r_h$ is the largest root of $f(r)=0$. To match the asymptotically Schwarzschild-AdS boundary condition,  functions $f$ and $\chi$ follow the asymptotic behavior
\begin{equation}\label{asymfchi1}
  f(r)=1-f_0/r^d+\cdots,~~\chi(r)=\mathcal{O}(1/r^{d+\delta}),~~\delta>0\,.
\end{equation}
This condition is stronger than asymptotically AdS conditions $f\rightarrow1$ and $\chi\rightarrow0$. The condition~\eqref{asymfchi1} will be satisfied in many holographic models under the so-called ``source-free condition'', such as the models in Refs.~\cite{Hartnoll:2008vx,Hartnoll:2008kx,Hartnoll:2009sz,Gubser:2008wv}. However, if one turns on the source terms, the condition~\eqref{asymfchi1} could be broken (see an example of scalar model in Ref.~\cite{Li:2020spf}).

We here allow a nonzero $\chi$ and so Eq.~\eqref{thermo-quantities} now becomes
\be
E = \fft{(d-1)\alpha}{4\pi} f_0 \,, \qquad  T = \fft{r_h^2e^{\chi(r_h)}f'(r_h)}{4\pi} \,, \qquad S = \alpha r_h^{d-1} \,, \label{thermo-quantities2}
\ee
and $E_0=0$ in this case. The energy-momentum tensor ${T^\mu}_\nu$ has a form
\begin{equation}\label{tmn1}
  {T^\mu}_{\nu}=\text{diag}[-\rho(r),p_r(r),p_T(r), p_T(r),\cdots,p_T(r)]\,.
\end{equation}
Einstein's equation gives the following three independent equations
\begin{eqnarray}
  f'&=&-\frac{16\pi\hat{\rho}}{(d-1)r}-\frac{df}r\,,\label{einstf}\\
  \chi'&=&\frac{8\pi}{(d-1)rf}(\rho+p_r)\,,\label{einstchi}\\
p_r'&=&\frac{(d-2)\hat{\rho}+2(d-1)\hat{p}_T-d\hat{p}_r}{2r}-\frac{8\pi(\hat{p}_r+\hat{\rho})\hat{p}_r}{(d-1)rf}\,.\label{einstpr}
\end{eqnarray}
Here
$$\hat{\rho}=\rho-\frac{(d-1)d}{16\pi}, ~~\hat{p}_r=p_r+\frac{(d-1)d}{16\pi},~~\hat{p}_T=p_T+\frac{(d-1)d}{16\pi}\,.$$
where $\rho, p_r$ and $p_T$ denote the matter contributions, and the hatted quantities include the cosmological constant (with the AdS radius set to $\ell=1$) into account.

Let us define a ``quasi-local mass'' for an equal-$r$ surface
\begin{equation}\label{defQ}
\begin{split}
  m(r)=\frac{r^{d+1}e^{-\chi}}{2d}(fe^{2\chi})'\,.
  \end{split}
\end{equation}
The boundary condition~\eqref{asymfchi1} also implies
\begin{equation}\label{minfity1}
   m(\infty)=f_0/2\,.
\end{equation}
and so Eq.~\eqref{thermo-quantities2} gives
\begin{equation}\label{teeeq1}
  (E-E_0)=(d-1)\alpha f_0/4\pi=(d-1)\alpha m(\infty)/2\pi
\end{equation}
Evaluating $m(r)$ at the horizon  gives us
\begin{equation}\label{minfity2}
   m(r_h)=\frac{r_h^{d+1}e^{\chi(r_h)}}{2d}f'(r_h)\Rightarrow TS=\alpha m(r_h)d/2\pi\,.
\end{equation}
One can verify by using Eqs.~\eqref{einstf}-\eqref{einstpr} that
\begin{equation}\label{dmdr}
  m'(r)=8\pi\frac{r^{d-1}\e^{\chi}}{d}(\hat{\rho}+\hat{p}_T)=8\pi\frac{r^{d-1}\e^{\chi}}{d}(\rho+p_T)\,.
\end{equation}
Under NEC, this means that $m(r_h)\leq m(\infty)$. Combining this with Eqs.~\eqref{teeeq1} and \eqref{minfity2} gives $(E-E_0)d\geq (d-1)TS$. This is a complete proof of bound~\eqref{energy bound} for static planar asymptotically Schwarzschild-AdS black branes satisfying the NEC, without assuming $g_{tt} g_{rr}=-1$. This independent proof also provides a nontrivial consistency check for the differential picture motivated by \eqref{ads-CQ-S}, although it does not by itself establish the local inequality~\eqref{ads-CQ-S}.

\end{document}